\documentclass[runningheads,a4paper]{llncs}
\usepackage{epsfig,shadow}
\usepackage{mdwmath}
\usepackage{mdwtab} 
\usepackage[linesnumbered,lined,boxed,commentsnumbered,ruled]{algorithm2e}
\usepackage{enumitem}
\usepackage{subfig}
\usepackage{tikz}
\usepackage{array}
\usepackage{multirow}
\usepackage{longtable}
\usepackage{amsmath}
\usepackage{amssymb}
\usepackage{rotate}
\usepackage{epsfig}
\usepackage{color}
\usepackage{graphicx}
\usepackage{psfrag}
\usepackage{multicol}
\usepackage{float}
\usepackage{url}
\urldef{\mailsa}\path|kissami.imad@lipn.univ-paris13.fr|
\newcommand{\keywords}[1]{\par\addvspace\baselineskip
\noindent\keywordname\enspace\ignorespaces#1}

\begin{document}

\mainmatter

\title{Towards Parallel CFD computation\\ for the ADAPT framework}

\author{Imad Kissami${}^\dagger{}^\ast$ \and Christophe Cerin$^\dagger$
  \and Fayssal Benkhaldoun$^\ast$ \and Gilles Scarella$^\ast$}
\authorrunning{ }
\titlerunning{ }

\institute{Universit\'e de Paris 13\\
$^\dagger$ LIPN, $^\ast$ LAGA,\\
99, avenue Jean-Baptiste Cl\'ement, 93430 Villetaneuse, France\\
\mailsa\\
}
\toctitle{Lecture Notes in Computer Science}
\tocauthor{Authors' Instructions}
\maketitle

\begin{abstract} 
  In order to run Computational Fluid Dynamics (CFD) codes on
  large scale infrastructures, parallel computing has to be used
  because of the computational intensive nature of the problems. In
  this paper we investigate the ADAPT platform where we couple flow
  Partial Differential Equations and a Poisson equation. This leads to a linear system
  which we solve using direct methods. The implementation deals with the MUMPS parallel
  multi-frontal direct solver and mesh partitioning methods using
  METIS to improve the performance of the framework. We also
  investigate, in this paper, how the mesh partitioning methods are
  able to optimize the mesh cell distribution for the ADAPT
  solver. The experience gained in this paper facilitates the move to
  the 3D version of ADAPT and the move to a Service Oriented view of
  ADAPT as future work.

\end{abstract}

\section{Introduction and context of the work}

In the last recent decades CFD (Computational Fluid Dynamics) used to play an important
role in industrial designs, environmental impact assessments and
academic studies. The aim of our work is to fully parallelize an
unstructured adaptive code for the simulation of 3D streamer
propagation in cold plasmas. As a matter of fact, the initial sequential version of
the Streamer code needs up to one month for running typical benchmarks.

The generation of streamer discharges is described by coupling
electrostatic to the motion of charged particles (electrons, positive
and negative ions). The electrostatic is represented by a Poisson
equation for the electric potential and the motion of particles is
described by a set of convection-diffusion-reaction equations.

We aim to parallelize both the linear solver issued from the Poisson
equation and the evolution equation using domain decomposition and
mesh adaptation at 'the same time'. To our knowledge this is the first
time that such a challenge is considered. Intuitively speaking,
separating the two steps may add delays because we need to synchronize
them and to align the execution time on the slowest processor.
Considering the two steps simultaneously offers the potential to
better overlap different computational steps.

Other authors have tackled the parallelization of either the linear
solvers or the evolution equation usually without mesh adaptation.  In
\cite{DBLP:journals/mcs/AssousSS11} the authors consider the
parallelization of a linear system of electromagnetic equation on non
adaptive unstructured mesh. Their time integration method leads to the
resolution of a linear system which requires a large memory capacity
for a single processor.

In \cite{Vecil20141703} the authors introduce a parallel code which was
written in C++ augmented with MPI primitives and the LIS (Linear
Iterative Solver) library. Several numerical experiments have been
done on a cluster of 2.40 GHz Intel Xeon, with 12 GB RAM, connected
with a Gigabit Ethernet switch.  The authors note that a classical run
of the sequential version of their code might easily exceed one month
of calculation. The improvement of the parallel version is due to the
parallelization of three parts of the code: the diagonalization of the
Schr\"{o}dinger matrix, advancing one step in the Newton–-Raphson
iteration, and the Runge-–Kutta integrator.

The authors in \cite{Notay2015237} focused only on the
parallelization of a linear solver related to the discretization of
self-adjoint elliptic partial differential equations and using
multigrid methods.

It should also be mentioned that authors in \cite{kumar2013high} have successfully
studied similar problems to those presented in this paper. The associated
linear systems have been solved using iterative Gmres
and CG solvers \cite{doi:10.1137/0907058}. One difference is that in our work we consider
direct methods based on LU decomposition using the MUMPS solver.

Moreover, the direct solution methods generally involve the use of
frontal algorithms in finite element applications. The advent of
multi—-frontal solvers has greatly increased the efficiency of direct
solvers for sparse systems. They make full use of high performance
software layers such as invoking level 3 Basic Linear Algebra
Subprograms (BLAS)
\cite{Blackford01anupdated,DBLP:journals/toms/BoisvertD02}
library. Thus the memory requirement is greatly reduced and the
computing speed greatly enhanced. Multi—-frontal solvers have been
successfully used both in the context of finite volume,
finite element methods and in power system simulations.

The disadvantage of using direct solvers is that the memory size
increases much more quickly than the problem size itself. To
circumvent this problem, out-of-core multi—-frontal solvers have been
developed which have the capability of storing objects of the
resolution on the disk during factorization. Another viable
alternative is to use direct solvers in a distributed computing
environment. MUMPS
\cite{DBLP:conf/para/AmestoyDLK00,DBLP:journals/pc/AmestoyGLP06} is
among the fastest parallel general sparse direct solvers available
under public domain.

In order to deal with complex geometries and fluid flows, a large
number of mesh cells should be used. Therefore the parallel computing
paradigm has to be introduced for coping with this large number of
mesh cells. In the parallel computing field, several factors, such as
the load balancing, the number of neighboring sub-domains and the halo
cells (cells which are at the boundary of sub-domains), affect the
performance. A well balanced load has the potential to reduce the
amount of waiting processors and the partitioning method can optimize
the distribution of the mesh cells across processors, thus it will
improve the performance of the parallel application too.

In this paper, we also use METIS
\cite{Karypis:1998:FHQ:305219.305248}, an open source mesh
partitioning software, and we incorporate it into the streamer code of
ADAPT \cite{FBJFKHJK}, to solve the evolution PDE, in order to study the impact of mesh
partitioning on the parallel version of ADAPT that we are currently
developing.

Moreover we carry out experiments using the parallel multi—-frontal direct
solver (MUMPS) with matrices extracted from the streamer code of the
ADAPT platform.  The general strategy is as follows. The linear system
of equations is evaluated on different processors corresponding to the
local grid assigned to the processor.  The right hand side vector is
assembled on the host processor and it is injected into the MUMPS
solver. At the last step of the MUMPS solver, the solution is assembled
centrally on the host processor. This solution is then broadcast to
all the processors. We discuss later the pro and cons of such a
strategy.

The organization of the paper is the following. In section
\ref{sec:deux} we introduce the aim of the ADAPT platform and its
positioning. In section \ref{sec:trois} we discuss about parallel
approaches to solve evolution equation coupled with Poisson equation,
show the strategy to parallelize our code and do some experiments with
METIS. In section \ref{sec:quatre} we analyze different parts of our
code in terms of speedup and efficiency and we provide numerical
results that show the efficiency  our work. 
Section \ref{conclusion} concludes the paper.

\section{The ADAPT framework}\label{sec:deux}

\subsection{Overview}



ADAPT \cite{FBJFKHJK} is an object oriented platform for
running numerical simulations with a dynamic mesh adaptation strategy
and coupling between finite element and finite volume methods.

ADAPT, as a CFD (Computational Fluid Dynamics) software package, has
been developed for realizing numerical simulation on an unstructured
and adaptative mesh for large scale CFD applications.

In this paper, in order to tackle the long running time necessary for
numerical simulation, we study the MUMPS and METIS toolkits and we
integrate them into the ADAPT framework. In fact we perform code
coupling between the two main steps of the studied problem.

The existing ADAPT implementation is a sequential C++ code for each
phenomenon, and the code requires a huge CPU time for executing a 3D
simulation. For example, the 3D streamer code may run up to 30 days
before returning results, for this reason we decided to parallelize
the code. This paper is an important step into this direction.



\subsection{Working environment}

To realize all the experiments (sequential and parallel) we worked on
the MAGI cluster÷footnote{http://www.univ-paris13.fr/calcul/wiki}
which is located at the University of Paris 13, and also on the Ada
cluster of Idris\footnote{http://www.idris.fr}, one of the national
computing facility in France.

\section{Parallel approach}\label{sec:trois}

In this paper, we parallelize the evolution equation coupled with the
Poisson equation, written as:
\begin{equation}
  \left\{
  \begin{aligned}
    \frac{\partial u}{\partial t} + F(V,u)=S, \\ \\
    \Delta P=b.
  \end{aligned}
  \right.
\end{equation}

given that $F(V,u)=div(u.\overrightarrow{V})-\Delta u$, $S=0$, and $V=\overrightarrow{\nabla}P$, the previous system gives :

\begin{equation}
  \left\{
  \begin{aligned}
    \frac{\partial u}{\partial t}+div(u.\overrightarrow{V})=\Delta u\\
    \Delta P=b.
  \end{aligned}
  \right.
\end{equation}

The first equation is discretized using the finite volume
method on an unstructured triangular mesh. The time-integration of the
transport equation is performed using an explicit scheme. The
discretized form of Poisson equation leads to a linear algebraic
system. We obtain the following set of equations:

\begin{equation}
  \left\{
  \begin{aligned}
    u_i^{n+1}=u_i^n-\frac{\Delta t}{\mu_i}\underbrace{\sum_{j=1}^{m}u_{ij}\overrightarrow{V}_{ij}\overrightarrow{n}_{ij}|\sigma_{ij}|}_{Rez\_conv}+
    \frac{\Delta t}{\mu_i}\underbrace{\sum_{j=1}^{m} \overrightarrow{\nabla}u_{ij}\overrightarrow{n}_{ij}|\sigma_{ij}|}_{Rez\_dissip}\\
    A.\overrightarrow{P}^{n}=\overrightarrow{b}^{n}\label{14}
  \end{aligned}
  \right.
\end{equation}

where m is the number of faces of volume $\mu_i$,
$\overrightarrow{n}_{ij}$ is the unit normal vector of the face
$\sigma_{ij}$ (face between volumes {$\mu_i$ and $\mu_j$) and
$|\sigma_{ij}|$ is its length.  Other variables denoted by subscript
ij represent variables on the face $\sigma_{ij}$.

A is a large sparse matrix, the coefficients of the matrix depend only
on the grid topology.

\paragraph{Computation of $u_{ij}$:} In figure \ref{fig:Computation of
  uij} we took the example when the mesh is splitted into two subdomains,
and we compute $u_{ij}$ on the face $\sigma_{ij}$. For a given face $\sigma_k$, suppose $T_i$ and $T_j$ are respectively the cells at the left and right of $\sigma_k$. Let's note $\sigma_k = \sigma_{ij}$
and $u_{k}=u_{ij}$. The algorithm \ref{algo:algorithm1} shows how the $u_{ij}$ is computed on face $\sigma_{ij}$.

\begin{figure}[t!]
  \begin{center}
    \includegraphics[width=3in]{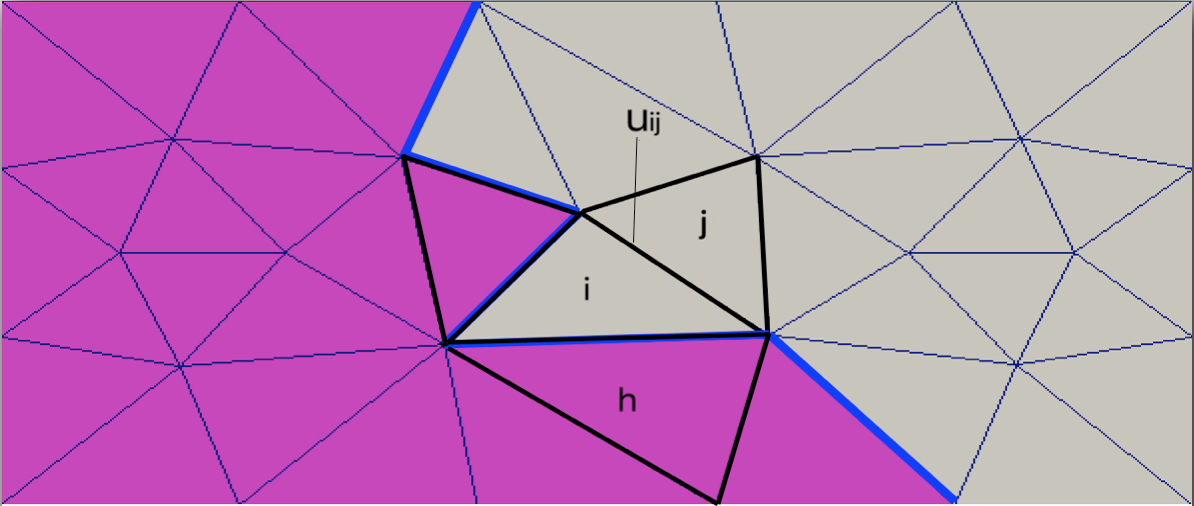}
  \end{center}
  \caption{Computation of $u_{ij}$ in mesh splitted into two subdomains}
  \label{fig:Computation of uij}
\end{figure}

\IncMargin{1em}
\begin{algorithm}[h!]
  F:number of faces\;
  \For{k:=1 \emph{\KwTo}  F}
      {
        \If{$dot(V_k.n_k) >= 0 $}
           {
          $u_k=u_i$\;
           }
           \Else{
             \If{$\sigma k $ is inner faces}{
               $u_k=u_j$\;
             }
             \ElseIf{$\sigma k $ is halo faces}{
               $u_k=u_h$;\tcc*[h]{$u_h:$ halo value sent by neighbor subdomain}
             }
           }
      }
      \caption{Compute $u_{ij}$ on face $ij$ }
      \label{algo:algorithm1}
\end{algorithm}

\paragraph{Computation of} $\overrightarrow{\nabla}u_{ij}$: In figure
\ref{fig:Computation of nablaij} we take the same mesh, and we
compute $\overrightarrow{\nabla}u_{ij}$ of the face $\sigma_{ij}$. The
algorithm \ref{algo:algorithm2} shows how the
$\overrightarrow{\nabla}u_{ij}$ is computed on volume $\mu_{ij}$. The
diamond cell in \ref{fig:Diamond cell} is constructed by connection of
centers of gravity $(i,j)$ of cells $T_i$, $T_j$ which shares the face
$\sigma_{ij}$ and its endpoints $A, B$.

\begin{figure}
  \begin{center}
    \includegraphics[width=3in]{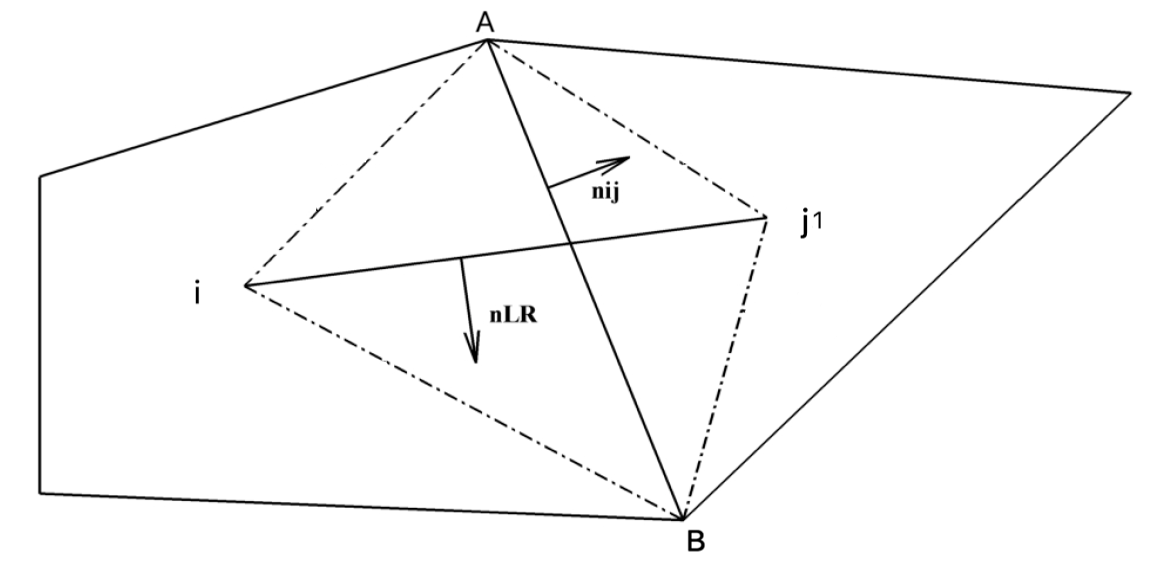}
  \end{center}
  \caption{Diamond cell in 2D}
  \label{fig:Diamond cell}
\end{figure}

\begin{equation}
\overrightarrow{\nabla}u_{ij}=\frac{1}{2\mu(D_{\sigma_{ij}})}[(u_A-u_B)\overrightarrow{n}_{LR}|\sigma_{LR}|+(u_j-u_i)\overrightarrow{n}_{ij}|\sigma_{ij}|]\\
\end{equation}

\IncMargin{1em}
\begin{algorithm}
  $u_{node}$:double\;
  $N$ :number of nodes\;
  $Alpha$ :weight coming from the least square method\;
  \For{n:=1 \emph{\KwTo} N }{
    \For{c:=1 \emph{\KwTo} inner cells around node n}{
      $u_{node}(n)+=Alpha * u_j(c)$\;
    }
    \For{m:=1 \emph{\KwTo} halo cells around node}{
      $u_{node}(n)+=Alpha * u_h(m)$\;
    }
  }
  \Return $u_{node}$\;
  \caption{Compute value at node ($u_{node}$) }
  \label{algo:algorithm2}
\end{algorithm}

\begin{figure}
  \begin{center}
    \includegraphics[width=3in]{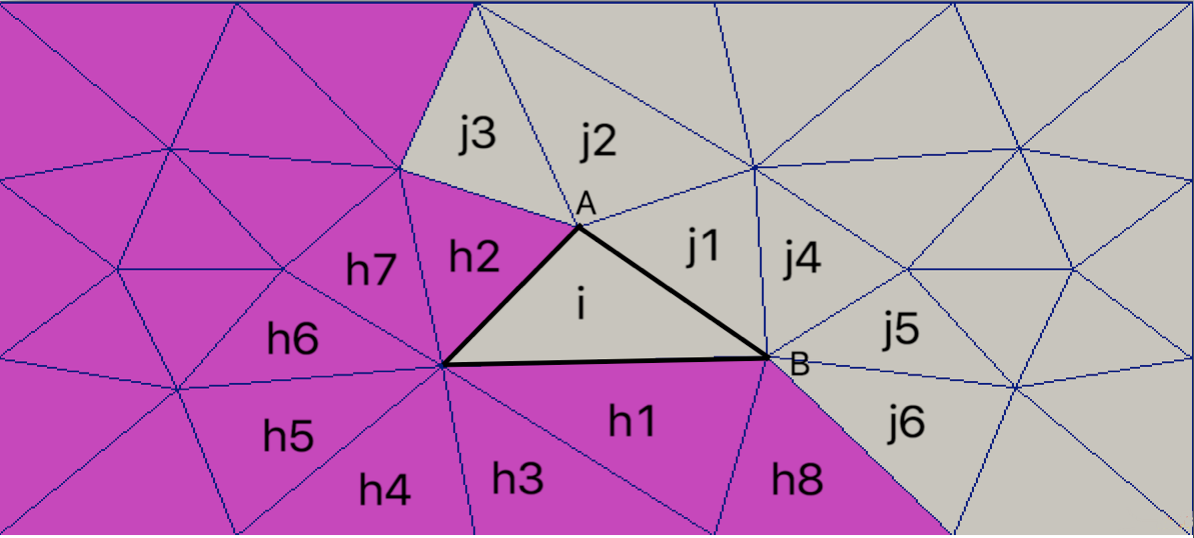}
  \end{center}
  \caption{Computation of $\nabla{u_{ij}}$ in mesh splitted into two subdomains}
  \label{fig:Computation of nablaij}
\end{figure}

\IncMargin{1em}
\begin{algorithm}
  F:number of faces\;
  $\overrightarrow{\nabla}u$:Vector2d\;
  \For{k:=1 \emph{\KwTo} F }
      {
        A:first node of face\;
        B:second face node\;
        i:center of gravity of cell $T_i$\;
        j:center of gravity of cell $T_j$\;
       $mes=\frac{1}{2\mu(D_{\sigma_{ij}})}$\;
        \If{Inner faces}{
          $\overrightarrow{\nabla}u(k)=mes * (u_{node}(A)-u_{node}(B))\overrightarrow{n}_{LR}|\sigma_{LR}| + (u_j-u_i)\overrightarrow{n}_{ij}|\sigma_{ij}|$\;
        }
        \ElseIf{Halo faces}{
          $\overrightarrow{\nabla}u(k)=mes * (u_{node}(A)-u_{node}(B))\overrightarrow{n}_{LR}|\sigma_{LR}| + (u_h-u_i)\overrightarrow{n}_{ih}|\sigma_{ih}|$\;        
        }
      }
      \Return $\overrightarrow{\nabla}u$\;
      \caption{Compute ${\nabla}u_{ij}$ on face $ij$ }
      \label{algo:algorithm3}
\end{algorithm}

\paragraph{Computation of the linear system:} The system is solved
directly by LU decomposition with an implementation for sparse
matrices. We use the \textbf{Intel MKL
  library}\footnote{https://software.intel.com/en-us/intel-mkl\#pid-3374-836}
solvers and \textbf{UMFPACK} \cite{DBLP:journals/toms/Davis04a}.

Note that in traditional approches, one splits the mesh to solve both
the evolution equation and the linear system in each sub-domain using
iterative methods with added communications. In our work we use the
same methodology in partitioning the mesh but the linear system is
solved with a direct method.


\paragraph{Parallelization:} For the parallelization of CFD
simulations, ADAPT employs the domain decomposition method. The 2D
unstructured mesh in Figure \ref{fig:maillage 2D} was decomposed into
eight subdomains using the METIS algorithm, these partitions have
approximatively the same size that we may consider as a good property
because the workload will be balanced on the homogenous processors of
our platform.

\begin{figure}
  \centering
  \includegraphics[width=3in]{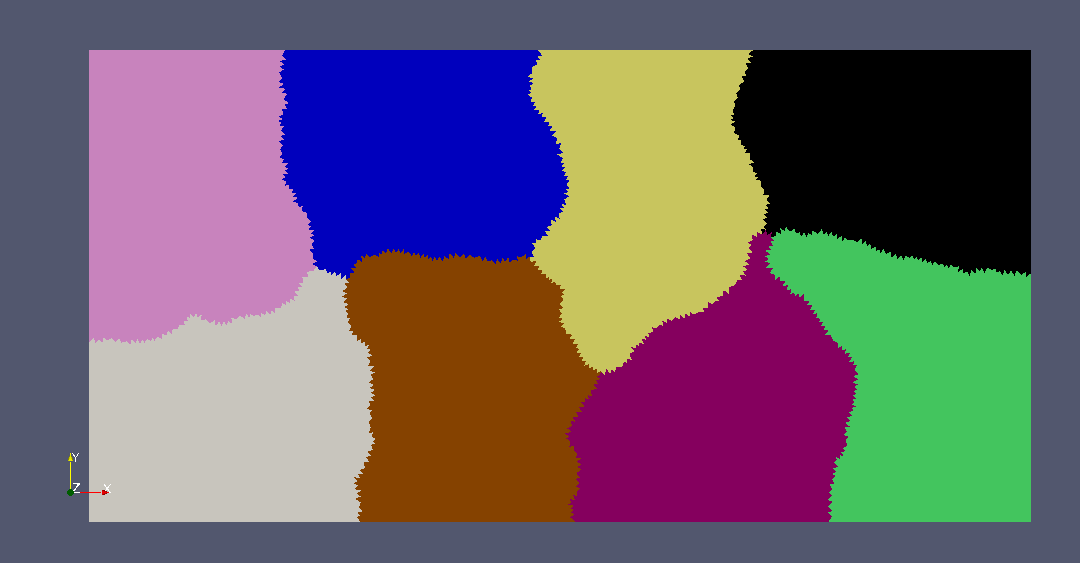}
  \caption{Decomposition of the computational domain into 8 subdomains using the METIS method }
  \label{fig:maillage 2D}
\end{figure}

Algorithm \ref{algo:algorithm} on page 10 provides with a piece of
pseudocode that shows how the parallelization is done in the streamer
code according to the coupling of evolution equation with Poisson
equation. We can see that most parts of the code are parallel ones
(line 19, line 9 to 11 and line 22 to 26) except reading and splitting
mesh in the beginning and between lines 14 and 15. Indeed, at this
step, we construct matrix A (eq. 2) that will be used to solve the
linear system in line 19. Matrix A is computed one time because it
depends only on the mesh.

Our contribution will serve in the future to tackle the 3D version of
streamer which includes the same type of equations, same strategy to
solve the linear system.  The difference is the structure of the
grid because we will work with Tetrahedra instead of triangles, which will
make the problem much more complicated due to big differences in handling
the mesh in the 3D case.

\IncMargin{1em}
\begin{algorithm}[h!]
  W:double\;
  \If(\tcc*[h]{for the master processor}){rank==0}{
    Read mesh data\;
    Split mesh with METIS\;
    Distribute mesh to all processors\;
  }
  W=0\;
  \For(\tcc*[h]{for each processor}){each rank}{
    Initialize conditions and create constants\;
    Send the informations of halos cells to neighbor subdomains\;
    Apply boundary conditions\;
  }
  \If{rank==0}{
    Construct matrix of linear system\;
    Split the matrix and send part of each processor\;
  }
  \For{each iteration}{
    \For(\tcc*[h]{for all processors}){all rank}{
      Solve linear system using MUMPS;
    }
    \For{each rank}{
      Send the informations of halos cells to neighbor subdomains\;
      Apply boundary conditions\;
      Compute fluxes of convection, diffusion and source term\;
      Update solution : $W^{n+1}=W^n+\Delta t * (rez\_conv + rez\_dissip + rez\_source)$\;
      Save results in parallel way using Paraview\;
    }
  }
  \caption{Algorithm of parallel ADAPT}
  \label{algo:algorithm}
\end{algorithm}

\section{Application to streamer equations}\label{sec:quatre}

When non-ionized or low ionized matter is exposed to high intensity electric field, non-equilibrium ionization processors (so-called
discharges or streamers) occur. Because of the reactive radicals they emit, streamers are used for the treatment of contaminated media like exhaust gasses,
polluted water or bio-gas.

\subsection{The governing equation}

The streamer consists of a convection-diffusion-reaction for the electron density, an ordinary differential equation for the positive ion density coupled by the Poisson's equation
for the electric potential. The model is given by the following equations:

\begin{eqnarray}
\frac{\partial n_e}{\partial t} +
div(n_e\overrightarrow{v_e} - De\overrightarrow{\nabla}n_e) = S_e ,\\ \label{1}
\ \ \frac{\partial n_i}{\partial t} =S_e ,\\ \label{2}
\Delta V= -\frac{e}{\epsilon}(n_i - n_e)\ , \label{3} \\ 
\ \ \vec E= -\overrightarrow{\nabla} V \ , \label{4}
\end{eqnarray}

Where $V$ is a potential of electric field E,  $\epsilon$ is the dielectric
constant, e the electron charge, $n_e$ and $n_i$ are the
number densities of electrons and positive ions, the drift velocity of
electrons is $v_e = v_e(E)$ and $D_e = D_e(E,v_e)$ is the
diffusion coefficient.  The source terms depend on the electron drift
velocity and the electron density $S_e=S_e(v_e,n_e)$.\\

\subsection{Numerical method}

The equations of the model are discretized using the finite volume method on an unstructured triangular mesh. The time-integration of the transport equations is performed using an explicit scheme.
The discretized form of Poisson’s equation consists of a linear algebraic system that is solved with a direct method at each time step during the time-integration. We approximate the equation for the
electron density (eq. 3) by the following finite volume method:
\begin{eqnarray}
\frac{\partial n_e}{\partial t}+ \frac{1}{\mu(T)}\ \oint_{\partial T}(n_e\overrightarrow{v_e}-D_e\overrightarrow{\nabla}n_e)\overrightarrow{n}\mathrm{d}s=S_e
\end{eqnarray}
where $\mu(T)$ is the volume of the cell T, $\overrightarrow{n} $ the
outward unit normal vector to the faces of the cell T.

The Poisson's equation (eq. 5) is discretized by a central type
approximation which leads to a system of linear equation, as follows:
\begin{eqnarray}
A.\overrightarrow{V}^{n+1}=\overrightarrow{b}^{n+1}
\end{eqnarray} 
$A$ is a matrix of coefficients, $\overrightarrow{V}$ is a vector of
unknowns (its dimension is equaled to the total number of cells) and
$\overrightarrow{b}$ is a vector of right hand side.

\subsection{Parallel results}

Table \ref{t:two} summarize our results. The speedup in this example
shows a good scalability of the present method for solving the
convection-diffusion equation coupled with Poisson equation
problems using mesh with 529240 cells. For practical applications, the computation time could be reduced from 49h54min (one computing core) down to 5min 
(1024 computing cores). This test is made on MAGI cluster at Paris13.


\begin{table}
  \centering
  \caption{Execution time (in s) of different parts of parallel 2D streamer code using mesh with 529240 cells}
\begin{tabular}{| c | c | c | c | c |}
  \hline
  Compute  & Total & Convection & Diffusion & Linear solver\\
  cores    &       &           &           &  \\
  \hline
  1& 49 h 54 min 48 s  & 02 h 51 min 04 s  & 13 h 06 min 00 s  & 33 h 57 min 44 s \\ 
  2& 25 h 06 min 27 s  & 01 h 22 min 57 s  & 06 h 41 min 02 s  & 17 h 02 min 27 s \\
  4& 12 h 34 min 35 s  & 00 h 42 min 07 s  & 03 h 22 min 04 s  & 08 h 30 min 24 s \\
  8& 06 h 27 min 18 s  & 00 h 22 min 13 s  & 01 h 46 min 26 s  & 04 h 18 min 38 s \\
  16& 03 h 39 min 45 s  & 00 h 12 min 37 s  & 01 h 01 min 40 s  & 02 h 25 min 26 s \\
  32& 01 h 50 min 50 s  & 00 h 08 min 03 s  & 00 h 29 min 17 s  & 01 h 13 min 29 s \\
  64& 01 h 01 min 41 s  & 00 h 03 min 59 s  & 00 h 17 min 05 s  & 00 h 40 min 36 s \\
  128& 00 h 32 min 44 s  & 00 h 01 min 52 s  & 00 h 08 min 22 s  & 00 h 22 min 29 s \\
  256& 00 h 18 min 16 s  & 00 h 01 min 02 s  & 00 h 04 min 34 s  & 00 h 12 min 39 s \\
  512& 00 h 10 min 07 s  & 00 h 00 min 33 s  & 00 h 02 min 52 s  & 00 h 06 min 42 s \\
  1024& 00 h 05 min 14 s  & 00 h 00 min 18 s  & 00 h 01 min 33 s  & 00 h 03 min 23 s \\
\hline
\end{tabular}
\label{t:two}
\end{table}

\paragraph{Scalability measurement:} There are two basic ways to
measure the scalability of parallel algorithms:
\begin{enumerate}[label=(\alph*)]
\item Strong scaling with a fixed problem size but the number of
  processing elements are increased.
\item Weak scaling with problem size and compute elements increasing
  concurrently.
\end{enumerate}


The speedup is defined as
$$sp = \frac{t_b}{t_N},$$
and the ideal speedup, $sp_{ideal}$, is naturally equivalent to the
number of compute cores. Therefore, the strong scaling efficiency can
also be calculated by
$$\frac{s_p}{sp_{ideal}}$$
which shows how far is the measured speedup from the ideal one.

In the this work, the strong scaling measurement is employed and the
scaling efficiency is calculated by:
$$\frac{t_bN}{t_NN_b} 100, \, N \ge N_b$$
where $t_b$ is the wall clock time of a base computation, $N_b$ is the
number of compute cores used for the base computation, $t_N$ is the
wall clock time of a computation with $N$ compute cores.

Figures \ref{fig:Speedup} and \ref{fig:efficiency} illustrate the
speedup and the strong scaling efficiency in the computations of our
illustrate example with $t_b$ the wall clock times by one compute core.
A more detailed injection into the data of this example provides
interesting information about the computational efficiency of the
individual parts of the parallel simulation.  The results of the
strong scaling tests, plotted in Fig \ref{fig:efficiency} depict that
the parallel speedup of the different parts increases up to 1024
computational cores, so our parallel strategy shows its efficiency.
\begin{figure}
  \begin{center}
    \includegraphics[width=4in]{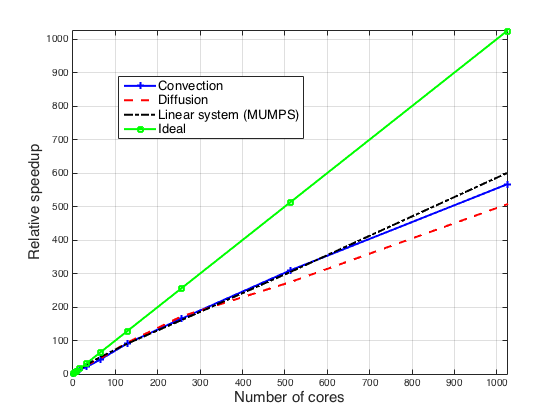}
  \end{center}
  \caption{Speedup of different parts of streamer code using mesh with 529240 cells}
  \label{fig:Speedup}
\end{figure}

\begin{figure}
  \centering
  \includegraphics[width=4in]{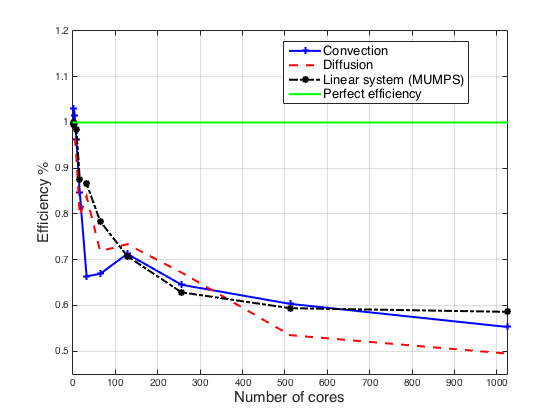}
  \caption{Efficiency of different parts of streamer code using mesh with 529240 cells}
  \label{fig:efficiency}
\end{figure}

\begin{figure}
  \centering
  \includegraphics[width=4in]{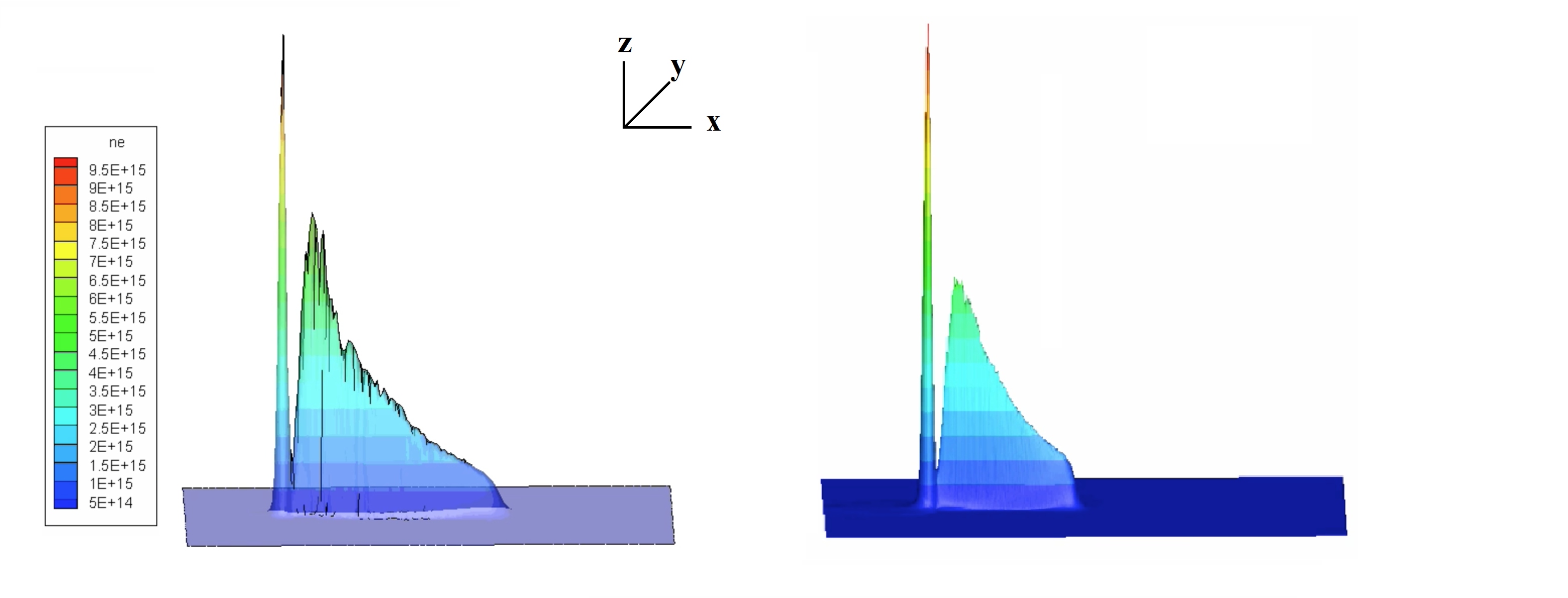}
  \caption{Plasma discharge in sequential code (left) and parallel one (right)}
  \label{fig:Plasma discharge in sequential code (left) and parallel one (right)}
\end{figure}

\subsection{Comparison between sequential and parallel code}

\begin{table}
  \centering
  \begin{tabular}{|c|c|c|}
    \hline
    & Compute & Execution\\
    & cores    & time\\
    \hline
    Mesh using 43674 elements &  1 & \color{red}04h30min\\
    \hline
    \multirow{2}{*}{Mesh using 529240 elements} & 1 & 150h15min\\
    & 64 & \color{red}03h59min\\
    \hline
  \end{tabular}
\caption{Comparison between sequential and parallel code}
\label{t:four}
\end{table}

In Figure \ref{fig:Plasma discharge in sequential code (left) and
  parallel one (right)} a comparison is made between sequential 2D
streamer code and parallel counterpart. We can also observe the
propagation of streamer in two cases: more the mesh is fine more the
results are accurate (without oscillations). In table \ref{t:four},
the parallel code takes the same running time (using 64 MPI processors)
as the sequential code but the input size is 12 times greater.

To conclude, our strategy although not classical shows to be very
efficient, and provides with relevant properties regarding performance.

\section{Conclusions and future work}\label{conclusion}

In this work, we have introduced an effective parallelization of the
ADAPT platform for CFD applications. We present the parallelization of
the convection-diffusion equation and the linear system. The
originality of this work is to solve linear system using a direct
method whereas existing studies mainly use iterative methods to solve
linear system in this kind of problems. The workflow is realized: (1)
using external tools such as METIS for mesh partitioning in order to
enable the computational load balance for the global assembly, (2)
using MPI to assure communication, (3) using MUMPS solver to solve the
system of linear equations. The current approach with MUMPS shows
significant advantages in terms of avoiding the problems of
pre-conditioners, and doesn't need a lot of iterations to converge to
the solution. In summary, the most important advantages of the
presented scheme are:
\begin{itemize}
\item Efficient memory usage distribution over computer nodes,
\item Good scalability to solve convection-diffusion equation,
\item Stable, fast and good scalability of linear solvers provided by MUMPS.
\end{itemize}

We observe that the campaign of numerical tests provides expected
results, meaning that our parallel implementation goes into the right
direction at the moment.

For the future, we plan to investigate in depth the mesh
partitioning problems and to conduct experiments with specific
tools. We may target PAMPA \cite{lachat:hal-00768916} and the question
is how such tool can impact the overall performance of ADAPT
software. The challenge is to mix mesh partitioning and dynamic mesh
adaptations. How to manage load balancing in this context? When
dealing with three dimensional problems, the question is even more
critical. At least, as explained in the discussion section our current
work will serve for solving 3D streamer problems and a service
oriented view of the ADAPT framework.

\section*{Acknowledgment}

The funding supports of this work is the EnCoMix AAP SPC project
(ANR-11-IDEX-05-02 Ref: SPC/JFG/2013‐031). The experiments conducted
in this work were done on the nodes B500 of University Paris 13
MAGI Cluster and available at http://www.univ-paris13.fr/calcul/wiki/

\bibliographystyle{unsrt}

\bibliography{article_ica3pp}

\end{document}